\documentclass[aps,prstab,twocolumn,groupedaddress]{revtex4-1}

\usepackage[alsoload=hep,binary-units,load=prefixed]{siunitx}
\usepackage{graphics,epsfig,color}

\newcommand{\be}{\begin{equation}}
\newcommand{\ee}{\end{equation}}

\newcommand{\beq}{\begin{eqnarray}}
\newcommand{\eeq}{\end{eqnarray}}

\begin{document}


\title{Phase modulation in storage-ring RF systems}


\author{D. Teytelman}
\affiliation{Dimtel, Inc., San Jose, CA}


\date{June 29, 2019}

\begin{abstract}

This paper presents a conceptual approach to phase modulation of the cavity
field in storage ring RF systems.  An implementation of the concept on
Dimtel low-level RF controllers is also presented.  The method is
illustrated with the test results from a cavity simulator, as well as an
electron storage ring KARA.

\end{abstract}

\pacs{}

\maketitle

\section{Introduction}
\label{sec:intro}
\medskip

Phase modulation of RF cavity field is commonly used in electron storage
rings to excite quadrupole oscillation of the beam.  This is typically done
to increase the effective bunch length and to improve the Touschek lifetime
\cite{Sakanaka:2000nz,Mitsuhashi:2001tp,Sommer:2018iaa}. Such modulation is normally generated by the low-level
RF (LLRF) system.  When the desired modulation frequency is within the
bandwidth of the cavity field control loop or loops, modulation is easily
achieved by applying the necessary signal to the feedback loop setpoint.  At
higher modulation frequencies, comparable or exceeding LLRF loop bandwidth,
the task is more complicated.  Firstly, the cavity field no longer precisely
follows the setpoint. Secondly, due to cavity detuning from the resonance,
upper and lower sidebands can be strongly asymmetric, converting pure phase
modulation at the setpoint to a mix of amplitude and phase modulations of
the cavity field.

This paper will describe a concept for creating precise phase modulation of
the cavity field and a practical implementation of this concept on Dimtel
LLRF controller (LLRF9). The derivation of the phase modulation method is
presented in Sec.~\ref{sec:method}.  Section~\ref{sec:meas} contains
the results of both laboratory and accelerator tests of this approach.

\section{Phase Modulation Concept}
\label{sec:method}

The goal is to produce phase modulation of the cavity field at a given
modulation frequency. The approach, presented in this paper, works as
follows:

\begin{enumerate}
\itemsep 0pt
\item Define the desired cavity field signal;
\item Measure the closed-loop transfer function from the station setpoint to
the cavity probe;
\item Calculate the setpoint signal that will produce the desired cavity
field under closed-loop operation;
\item Compute the relevant configuration parameters to generate the setpoint
waveform.
\end{enumerate}

\subsection{Desired Cavity Field Signal}

For modulation amplitude $\beta$ (in radians) and frequency $\omega_m$ we
have:

\begin{eqnarray*}
V_c(t) = V_\mathrm{C0} \cos(\omega_\mathrm{rf} t + \beta \sin(\omega_m t))
\end{eqnarray*}
where $V_0$ is the cavity field amplitude setpoint.
For small modulation amplitudes, we can drop the sidebands at the harmonics
of the modulation frequency, leaving the following expansion:

\beq
V_c(t) &=& V_\mathrm{C0} J_0(\beta)\cos(\omega_\mathrm{rf}t) \nonumber \\
       &+&  V_\mathrm{C0} J_1(\beta)\cos((\omega_\mathrm{rf}+\omega_m)t)\nonumber \\
       &-&  V_\mathrm{C0} J_1(\beta)\cos((\omega_\mathrm{rf}-\omega_m)t)
\label{eq:vc}
\eeq
where $J_0$ and $J_1$ are Bessel functions of the first kind.

\subsection{Transfer Function}

Let $H_\mathrm{cl}(\omega)$ be the closed loop transfer function between the
station setpoint and the cavity field probe:

\be
V_c(\omega) = H_\mathrm{cl}(\omega) V_s(\omega)
\label{eq:tf}
\ee
where $V_s$ is the setpoint voltage.  Using the network analyzer, integrated
in LLRF9, we measure $H_\mathrm{cl}(\omega)$ at three frequencies:
$\omega_\mathrm{rf}-\omega_m$, $\omega_\mathrm{rf}$\footnote{Transfer
function measurement at the RF frequency is difficult due to the large
interfering signal.  In practice, this measurement is approximated by an
average of the transfer function measurements made at small (3~\si{\hertz})
symmetric offsets around the RF frequency}, and
$\omega_\mathrm{rf}+\omega_m$.  Let $H_L$ and $H_U$ be normalized complex
gains at lower and upper sidebands respectively:

\begin{eqnarray*}
H_L &=&
\frac{H_\mathrm{cl}(\omega_\mathrm{rf}-\omega_m)}
{H_\mathrm{cl}(\omega_\mathrm{rf})}\\
H_U &=&
\frac{H_\mathrm{cl}(\omega_\mathrm{rf}+\omega_m)}
{H_\mathrm{cl}(\omega_\mathrm{rf})}
\end{eqnarray*}

In theory, closed-loop response at RF frequency should be unity for the loop
topology where feedback error signal is the difference of the setpoint and
cavity field probe signals. Normalization allows us to avoid shifting
cavity setpoint at the RF in response to transfer function measurement
errors.

\subsection{Setpoint Waveform}

Using Equations~\ref{eq:vc}--\ref{eq:tf}, we calculate the desired setpoint
voltage:

\beq
V_s(t) &=& V_\mathrm{C0} J_0(\beta)\cos(\omega_\mathrm{rf}t)\nonumber\\
       &+& \frac{V_\mathrm{C0}
J_1(\beta)}{|H_U|}\cos((\omega_\mathrm{rf}+\omega_m)t-\angle H_U)\nonumber\\
       &-& \frac{V_\mathrm{C0}
J_1(\beta)}{|H_L|}\cos((\omega_\mathrm{rf}-\omega_m)t-\angle H_L)
\label{eq:setpt}
\eeq

If one had access to an arbitrary waveform generator as the source of the
setpoint signal, our task would be complete --- just applying the waveform
described by Eq.~\ref{eq:setpt} to the system setpoint is sufficient. In
practice, one more step is needed to map this waveform to the modulation
capabilities of a physical low-level RF system.

\subsection{Mapping to LLRF9}

Setpoint signal generation in LLRF9 supports simultaneous amplitude and
phase modulation profiles \cite{dimtel:llrf9}. To generate the waveform in
Eq.~\ref{eq:setpt}, we need to express the setpoint profile in terms of four
parameters: voltage setpoint $V_0$, amplitude modulation magnitude $a_a$,
phase modulation magnitude $\beta_p$, and relative phase between amplitude and
phase modulation waveforms $\Delta\phi_\mathrm{ap}$.

Equation~\ref{eq:setpt} defines the setpoint waveform as a sum of three
terms --- RF carrier and two sidebands. Let's use the following definitions:

\begin{eqnarray*}
a_U & = & J_1(\beta) |H_U|\\
a_L & = & -J_1(\beta) |H_L|\\
\phi_U & = & -\angle H_U\\
\phi_L & = & -\angle H_L\\
\omega_U & = & \omega_\mathrm{rf} + \omega_m\\
\omega_L & = & \omega_\mathrm{rf} - \omega_m
\end{eqnarray*}
to shorten that equation to:

\beq
V_s(t)  &=&  V_\mathrm{C0} J_0(\beta)\cos{\omega_\mathrm{rf}t}\nonumber\\
        &+& a_U \cos(\omega_U t + \phi_U) + a_L \cos(\omega_L t + \phi_L)
\label{eq:setpt_short}
\eeq

Setpoint signal with amplitude and phase modulation is given by

\be
V_s(t) = V_0 (1 + a_a\cos(\omega_m t+\phi_a))\cos(\omega_\mathrm{rf}t + 
\beta_p \sin(\omega_m t + \phi_p))
\label{eq:setpt_llrf9}
\ee

Expanding Eq.~\ref{eq:setpt_llrf9} and discarding sidebands at the harmonics of
$\omega_m$ as well as the terms proportional to $a_a J_1(\beta_p)$, then
matching the coefficients of sine and cosine terms of three spectral lines
to those in Eq.~\ref{eq:setpt_short}, we get:

\beq
\phi_a & = & \tan^{-1}\frac{a_U \sin\phi_U - a_L\sin\phi_L}
{a_U\cos\phi_U+a_L\cos\phi_L}\\
\phi_p & = & \tan^{-1}\frac{a_U \sin\phi_U + a_L\sin\phi_L}
{a_U\cos\phi_U-a_L\cos\phi_L}\\
a_a &=& \frac{a_U\cos\phi_U+a_L\cos\phi_L}{V_\mathrm{C0} J_0(\beta)
\cos\phi_a}\\
\frac{J_1(\beta_p)}{J_0(\beta_p)} &=& \frac{a_U\cos\phi_U-a_L\cos\phi_L}
{2 V_\mathrm{C0} J_0(\beta) \cos\phi_p}
\label{eq:ph_bessel}
\eeq
Numerical solver is then used to compute modulation argument $\beta_p$ from
the ratio of Bessel functions in Eq.~\ref{eq:ph_bessel}.  Finally,

\begin{eqnarray*}
V_0 = \frac{V_\mathrm{C0} J_0(\beta)}{J_0(\beta_p)}
\end{eqnarray*}

\section{Measurements}
\label{sec:meas}

\subsection{Bench Measurements}

\begin{figure}
\begin{center}
\epsfig{file=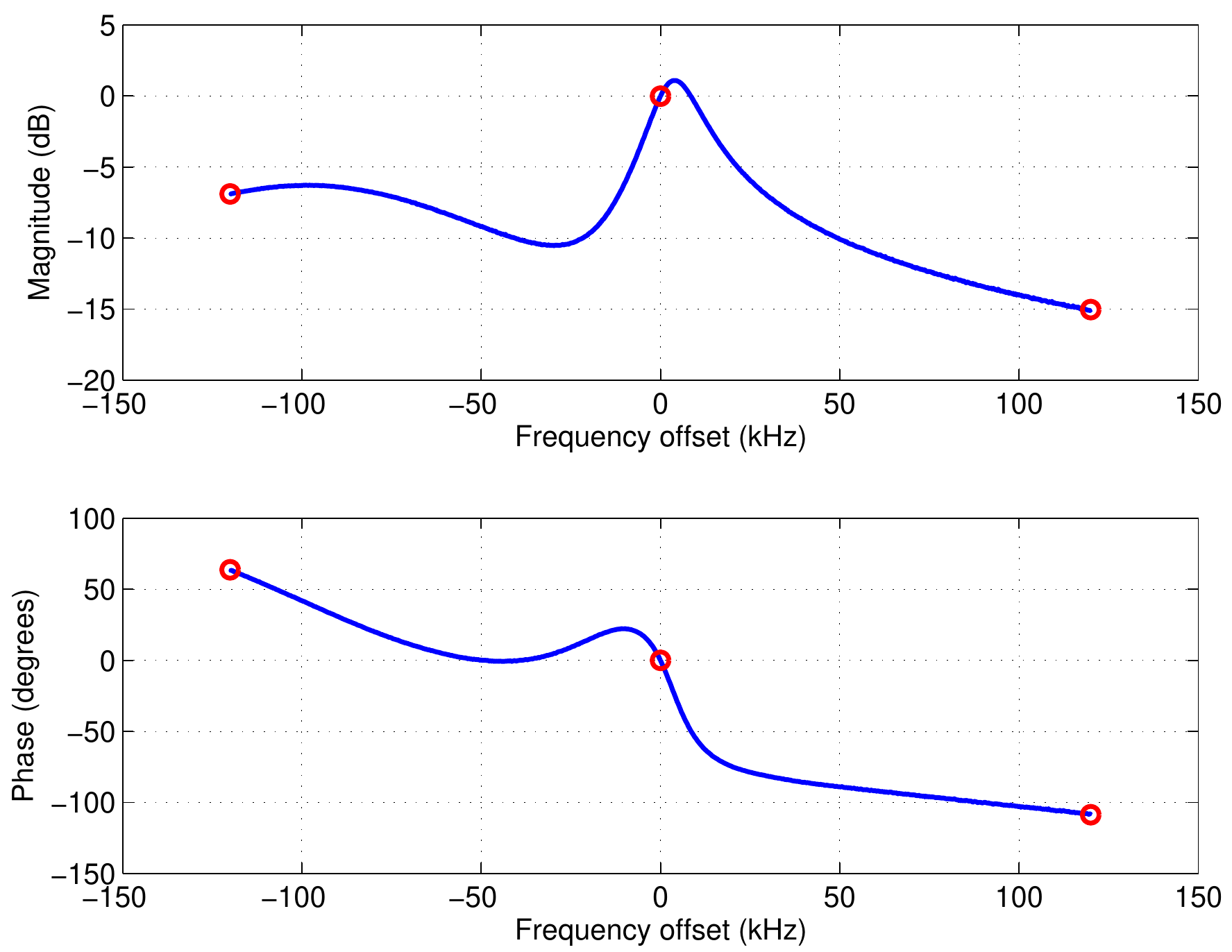,width=\columnwidth}
\end{center}
\caption{Transfer function measurement with a detuned cavity
filter. Red circles mark three values of $H(\omega)$ used to calculate the
modulation.}
\label{fig:TF}
\end{figure}

Bench measurements, presented in this section, were collected with a
cavity filter ($Q=7600$). The filter was tuned 80~\si{}{\kilo\hertz} below
the RF frequency, simulating typical situation in a storage ring and both
proportional and integral feedback loops were closed.
From the closed-loop transfer function measurement shown in
Figure~\ref{fig:TF}, setpoint parameters to get 1\si{\degree}
cavity field phase modulation at 120~\si{\kilo\hertz} were computed:

\begin{eqnarray*}
a_a = -0.039\\
\phi_a = -0.89\\
\beta_p = -0.065\\
\phi_p = -1.46\\
V_0 = 1.001V_\mathrm{C0}
\end{eqnarray*}

\begin{figure}
\begin{center}
\epsfig{file=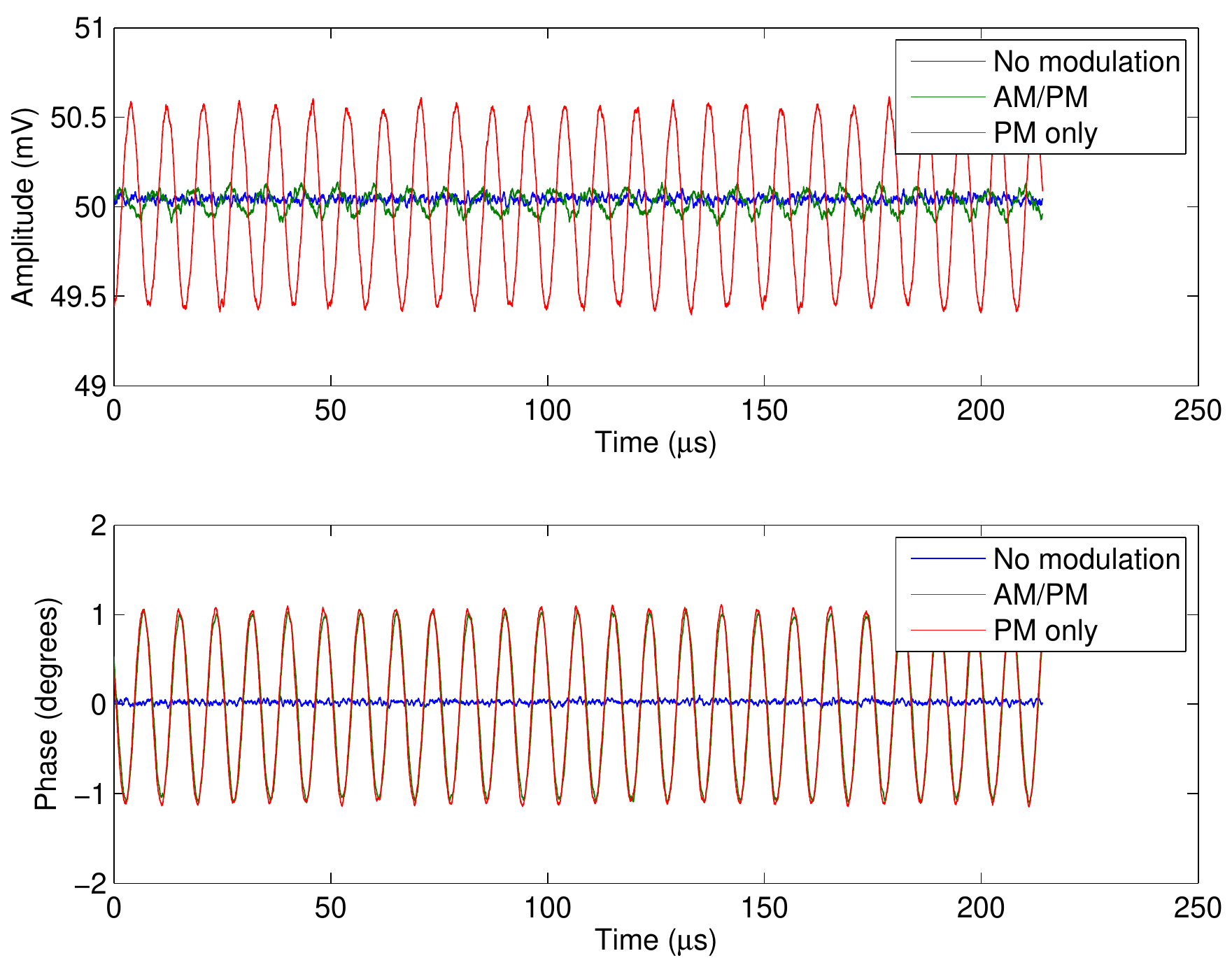,width=\columnwidth}
\end{center}
\caption{Cavity probe amplitude and phase for unmodulated (blue), fully
modulated (green), and phase-modulated (red) setpoint signals.}
\label{fig:modplot}
\end{figure}

Cavity output was then measured under three conditions: no modulation, full
modulation, as computed above, and phase modulation only.  Results are
presented in Fig.~\ref{fig:modplot}.  Standard deviation of the amplitude
rises from $3.4\times10^{-4}$ without modulation to $1.1\times10^{-3}$ with
combined amplitude and phase modulation of the setpoint.  Modulation leakage
into field amplitude is due to the approximations in the parameter
derivation as well as the errors in the transfer function measurement. When
the amplitude modulation term is removed, amplitude standard deviation rises
to $8\times10^{-3}$.

Setpoint phase modulation computed in this configuration is 3.7\si{\degree},
reflecting the magnitude of the transfer function. Cavity phase modulation
is 1.04\si{\degree} --- close to the design target.

\subsection{Accelerator Measurements}

Phase modulation technique was tested at KARA storage ring in Karlsruhe,
Germany. Main parameters of KARA are summarized in Table~\ref{table:params}.

\begin{table}[!htbp]
\caption{\bf KARA parameters} \label{table:params}
\begin{center}
\begin{tabular}{|l|c|}
\hline
Parameter & Value \\
\hline \hline
Injection energy, GeV            & 0.5 \\
Operating energy, GeV            & 2.5 \\
Beam current, \si{\milli\ampere} & 180 \\
Circumference, \si{\meter}       & 110 \\
Harmonic number $h$              & 184 \\
Synchrotron tune, $Q_s$          &  0.013 \\
\hline
\end{tabular}
\end{center}
\end{table}

KARA RF system consists of two stations, each with one klystron driving two
cavities. Phase modulation of 2\si{\degree} at 67~\si{\kilo\hertz} was
applied to one of two stations (Sector 4). Machine was operating at
2.5~\si{\giga\electronvolt} with insertion device gaps closed and beam
current of 30.9~\si{\milli\ampere}. 

\begin{figure}
\begin{center}
\epsfig{file=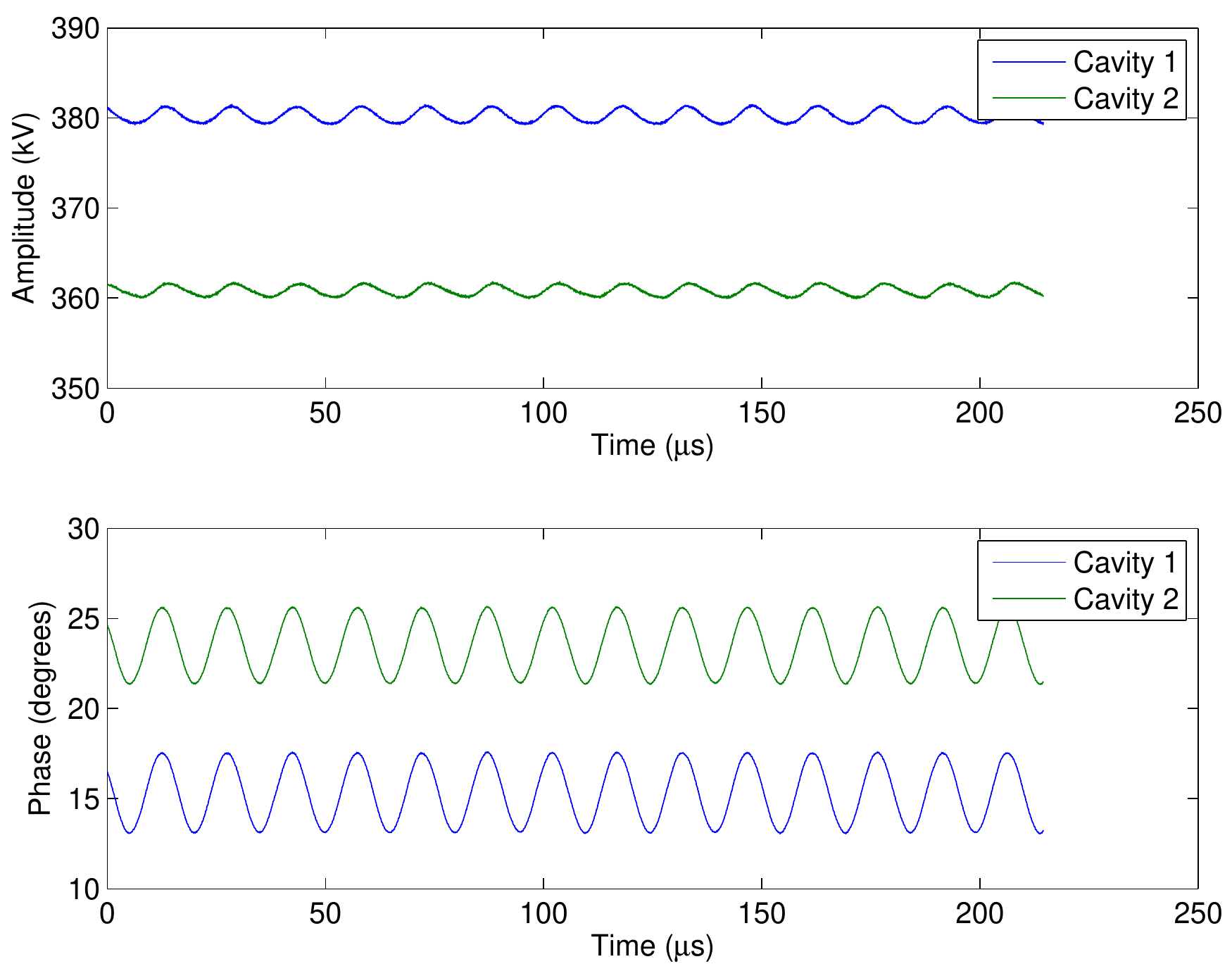,width=\columnwidth}
\end{center}
\caption{Amplitude and phase plots for cavities 1 and 2 in KARA RF station
S4.}
\label{fig:kara}
\end{figure}

Amplitude and phase signals in cavities 1 and 2 are shown in
Fig.~\ref{fig:kara}. The difference in average amplitudes and phases between
the two cavities is due to the station configuration imperfections. Phase
and amplitude modulation levels are 2.2\si{\degree}/0.18\% and
2.1\si{\degree}/0.15\% in cavities 1
and 2 respectively. Most likely reason for slightly elevated phase
modulation levels in the cavity signals is the imperfect tuning of the
vector combiner, with cavity signals being added at an angle. In this loop
configuration, there is no independent control of modulation levels in
individual cavities.

\section{Conclusions}

This works presents a method of generating phase modulation in an RF cavity
under beam loading.  The method seamlessly accounts for closed-loop response
of the RF station, including amplitude roll-off and asymmetry.  The method
was demonstrated on the bench as well as in a real accelerator with beam. 
In all cases, low level of amplitude modulation was seen.  The levels of
phase modulation in operation were in good agreement with the desired
setting.


\bibliography{llrf_phase_mod}

\end{document}